\documentstyle[preprint,aps,floats,epsf]{revtex}

\tightenlines 

\catcode`@=11
\def\references{%
\ifpreprintsty
\bigskip\bigskip
\hbox to\hsize{\hss\large \refname\hss}%
\else
\vskip24pt
\hrule width\hsize\relax
\vskip 1.6cm
\fi
\list{\@biblabel{\arabic{enumiv}}}%
{\labelwidth\WidestRefLabelThusFar  \labelsep4pt %
\leftmargin\labelwidth %
\advance\leftmargin\labelsep %
\ifdim\baselinestretch pt>1 pt %
\parsep  4pt\relax %
\else %
\parsep  0pt\relax %
\fi
\itemsep\parsep %
\usecounter{enumiv}%
\let\p@enumiv\@empty
\def\theenumiv{\arabic{enumiv}}%
}%
\let\newblock\relax %
\sloppy\clubpenalty4000\widowpenalty4000
\sfcode`\.=1000\relax
\ifpreprintsty\else\small\fi
}
\catcode`@=12

\begin{document}


\preprint{
\font\fortssbx=cmssbx10 scaled \magstep2
\hbox to \hsize{
\hbox{\fortssbx University of Wisconsin - Madison}
\hfill$\vcenter{\hbox{\bf MADPH-97-992}
                \hbox{hep-ph/9704403}}$ }}

\title{\vspace{.5in}
Relic Density of Neutralino Dark Matter \\ in Supergravity Models}
 
\author{V. Barger and Chung Kao}

\address{
Department of Physics, University of Wisconsin, 
Madison, WI 53706}

\maketitle

\thispagestyle{empty}
 
\bigskip

\begin{abstract}

We calculate the relic density ($\Omega_{\chi^0_1} h^2$) 
of neutralino dark matter in supergravity models 
with radiative electroweak symmetry breaking,  
for values of the Higgs sector parameter $\tan\beta \equiv v_2/v_1$ 
between the two infrared fixed points of the top quark Yukawa coupling, 
$\tan\beta \simeq 1.8$ and $\tan\beta \simeq 56$. 
For $\tan\beta \alt 10$, 
the CP-odd Higgs pseudoscalar ($A^0$) 
and the heavier CP-even Higgs scalar ($H^0$) 
are much heavier than the lightest neutralino ($\chi^0_1$), 
and the annihilation cross section of $\chi^0_1$ pairs 
is significantly enhanced only in the vicinity 
of the lighter CP-even Higgs boson ($h^0$) and the $Z$ boson poles.
For $\tan\beta \agt 40$, 
$m_A$ and $m_H$ become comparable to $2m_{\chi^0_1}$, 
and the neutralino annihilation cross section 
can be significantly enhanced by the poles of the $H^0$ and the $A^0$ as well. 
For a cosmologically interesting relic density, 
$0.1 \alt \Omega_{\chi^0_1} h^2 \alt 0.5$, 
we find that the supergravity parameter space 
is tightly constrained.
The mass spectra for supersymmetric particles 
are presented for unification mass parameters 
in the cosmologically interesting region.

\end{abstract}

%

\section{Introduction}

A supersymmetry (SUSY) between fermions and bosons 
provides a natural explanation 
of the Higgs electroweak symmetry breaking (EWSB) mechanism 
in the framework of a grand unified theory (GUT).  
The evolution of renormalization group equations (RGEs) \cite{RGE} 
with the particle content 
of the minimal supersymmetric standard model (MSSM) \cite{MSSM,Xerxes} 
is consistent with a grand unified scale 
at $M_{\rm GUT} \sim 2\times 10^{16}$ GeV 
and an effective SUSY mass scale in the range 
$M_Z < M_{\rm SUSY} \alt 1$ TeV. 
With a large top quark Yukawa coupling to a Higgs boson at the GUT scale, 
needed for $b-\tau$ Yukawa unification \cite{btau,BBO}, 
radiative corrections can drive the corresponding Higgs boson mass squared 
parameter negative, spontaneously breaking the electroweak symmetry 
and naturally explaining the origin of the electroweak scale.
In the minimal supersymmetric GUT 
with a large top quark Yukawa coupling ($Y_t$), 
there is an infrared fixed point (IRFP) \cite{BBO,IRFP1,IRFP2,IRFP3} 
at low $\tan\beta$, and the top quark mass is correspondingly predicted 
to be $m_t = 200 \; {\rm GeV} \sin\beta$ \cite{BBO}. 
At high $\tan\beta$, another quasi-IRFP solution ($dY_t/dt \simeq 0$) exists 
at $\tan\beta \sim 60$ \cite{BBO,IRFP2}. 
For $m_t = 175$ GeV, the two IRFPs of top quark Yukawa coupling 
appear at $\tan\beta \simeq 1.8$ and $\tan\beta \simeq 56$ \cite{BBO}.

Supersymmetry extends the Poincar\'e symmetry to new dimensions 
of space and time and local gauge invariance with SUSY includes gravity. 
In supergravity (SUGRA) models \cite{SUGRA}, 
it is assumed that supersymmetry is broken in a hidden sector 
with SUSY breaking communicated to the observable sector 
through gravitational interactions, 
leading naturally but not necessarily \cite{Non-universal} 
to a common scalar mass ($m_0$), a common gaugino mass ($m_{1/2}$), 
a common trilinear coupling ($A_0$) and a common bilinear coupling ($B_0$) 
at the GUT scale.
Through minimization of the Higgs potential, the $B$ parameter 
and magnitude of the superpotential Higgs mixing parameter $\mu$ 
are related to  the ratio of vacuum expectation values (VEVs) of Higgs fields 
($\tan\beta \equiv v_2/v_1$), and the mass of the $Z$ boson ($M_Z$).
The SUSY particle masses and couplings at the weak scale 
can be predicted by the evolution of RGEs 
from the unification scale \cite{BBO,SUGRA2}. 

In our analysis, we evaluate SUSY mass spectra and couplings 
in the minimal supergravity model 
with four parameters: $m_0$, $m_{1/2}$, $A_0$ and $\tan\beta$; 
and the sign of the Higgs mixing parameter $\mu$. 
The neutralino relic density is only slightly affected 
by a change in the $A_0$ parameter 
since  $A_0$ mainly affects the masses of third generation sfermions. 
Therefore, for simplicity we take $A_0 = 0$ in most of our calculations. 

The mass matrix of the neutralinos in the weak eigenstates 
($\tilde{B}$, $\tilde{W}_3$, $\tilde{H}_1$, $\tilde{H}_2$) 
has the following form \cite{BBO}
\begin{equation}
M_N=\left( \begin{array}{c@{\quad}c@{\quad}c@{\quad}c}
M_1 & 0 & -M_Z\cos \beta \sin \theta _W^{} & M_Z\sin \beta \sin \theta _W^{}
\\
0 & M_2 & M_Z\cos \beta \cos \theta _W^{} & -M_Z\sin \beta \cos \theta _W^{}
\\
-M_Z\cos \beta \sin \theta _W^{} & M_Z\cos \beta \cos \theta _W^{} & 0 & \mu
\\
M_Z\sin \beta \sin \theta _W^{} & -M_Z\sin \beta \cos \theta _W^{} & \mu & 0
\end{array} \right)\;.
\label{eq:xino}
\end{equation}
This matrix is symmetric and can be diagonalized by a single matrix 
to obtain the mass eigenvalues and eigenvectors. 
The form of Eq.~(\ref{eq:xino}) establishes our sign convention for $\mu$. 

Recent measurements of the $b \to s\gamma$ decay rate 
by the CLEO \cite{CLEO} and LEP collaborations \cite{LEP} 
place constraints on the parameter space 
of the minimal supergravity model \cite{bsg}. 
It was found that $b \to s\gamma$ excludes 
most of the minimal supergravity (mSUGRA) parameter space 
when $\tan\beta$ is large and $\mu > 0$ \cite{bsg}. 
In this article, we present results for both $\mu > 0$ and $\mu < 0$. 

We calculate the masses and couplings in the Higgs sector 
with one loop corrections from both the top and the bottom Yukawa interactions 
in the RGE-improved one-loop effective potential \cite{Higgs} 
at the scale $Q = \sqrt{m_{\tilde{t}_L}m_{\tilde{t}_R}}$. 
With this scale choice, the numerical value 
of the CP-odd Higgs boson mass ($m_A$)  at large $\tan\beta$ \cite{Scale} 
is relatively insensitive to the exact scale and the loop corrections to $M_A$ are small compared to the tree level contribution.
In addition, when this high scale is used, 
the RGE improved one-loop corrections approximately 
reproduce the dominant two loop perturvative calculation 
of the mass of the lighter CP-even Higgs scalar ($m_h$).  
Our numerical values of $m_h$ are very close to the results of 
Ref.~\cite{Two-Loop} where somewhat different scales higher than $M_Z$ 
have been adopted in evaluating the effective potential.
 
The matter density of the Universe $\rho$ is described 
in terms of a relative density $\Omega = \rho/\rho_c$ with 
$\rho_c = 3H_0^2/8\pi G_N \simeq 1.88 \times 10^{-29} h^2 \; {\rm g/cm^3}$ 
the critical density to close the Universe. 
Here, $H_0$ is the Hubble constant,
$h = H_0/( 100 \; {\rm km}\; {\rm sec}^{-1}\; {\rm Mpc}^{-1})$, 
and $G_N$ is Newton's gravitational constant.

There is compelling evidence from astronomical observations 
for the existence of dark matter \cite{DM0}.
Studies on clusters of galaxies and large scale structure 
suggest that the matter density in the Universe 
should be at least $20\%$ of the critical density ($\Omega_M \agt 0.2$) 
\cite{OmegaM}, but the big-bang nucleosynthesis 
and the measured primordial abundance of helium, deuterium 
and lithium constrain the baryonic density 
to $0.01 \alt \Omega_b h^2 \alt 0.03$ \cite{Baryon}. 
The anisotropy in the cosmic microwave background radiation 
measured by the Cosmic Background Explorer (COBE) 
suggests that at least $60\%$ of the dark matter 
should be cold (nonrelativistic) \cite{COBE}. 
In supersymmetric theories with a conserved $R$-parity\footnote{
$R = +1$ for the SM particles and Higgs bosons 
and $R = -1$ for their superpartners.}, 
the lightest supersymmetric particle (LSP) 
cannot decay into normal particles and the LSP is an attractive candidate
for cosmological dark matter \cite{Rocky,SUSYDM}. 
In most of the supergravity parameter space, 
the lightest neutralino ($\chi^0_1$) is the LSP \cite{BBO,SUGRA2}. 

Inflationary models usually require $\Omega = 1$ for the Universe 
\cite{Rocky}, although models with $\Omega < 1$ have recently
been proposed \cite{Inflation}. 
Recent measurements on the Hubble constant 
are converging to $h \simeq 0.6 - 0.7$ 
\cite{Hubble}.
Therefore, we conservatively consider the cosmologically interesting region 
for $\Omega_{\chi^0_1}$ to be
\begin{equation}
0.1 \alt \Omega_{\chi^0_1} h^2 \alt 0.5.
\label{eq:ohs}
\end{equation}

Following the classic papers of Zel'dovich, Chiu, and 
Lee and Weinberg \cite{Relic}, 
many studies of the neutralino relic density in supergravity models 
have been made with universal \cite{SUGRADM,Manuel,BWR,Howie}  
and nonuniversal \cite{Nonuniversal} soft breaking masses. 
Recent calculations have taken into account threshold effects 
and integration over Breit-Wigner poles \cite{Kim91,Gondolo}.
Previous analyses have focused on the small to moderate $\tan\beta$ range, 
$\tan\beta \alt 25$, but large $\tan\beta$ is equally of interest. 
Some theoretical models of fermion masses 
require a large value $\tan\beta$ \cite{tanb}.

In this article, 
we make more complete calculations of the neutralino relic density 
over the full range of $\tan\beta$ and SUGRA parameter space 
with universal GUT scale boundary conditions.
We employ the calculational methods of Ref.~\cite{Howie}, 
with some significant improvements. 
The neutralino annihilation cross section 
is evaluated with helicity amplitude techniques 
to eliminate any uncertainty from a neutralino velocity expansion. 
The neutralino relic density is calculated 
with relativistic Boltzmann averaging, 
neutralino annihilation threshold effects 
and Breit-Wigner poles. 
We introduce a transformation 
to improve the numerical integration over Breit-Wigner poles 
and use appropriate extension of the RGEs for large $\tan\beta$ solutions.
We determine the constraints on $m_{1/2}$ and $\tan\beta$ 
implied by Eq. (\ref{eq:ohs}). 
We also impose the constraints from the chargino search 
at LEP 2 \cite{ALEPH}. 
The effects of the common trilinear coupling ($A_0$) 
on $A_t$, $m_{\tilde{t}_1}$ and the neutralino relic density 
$\Omega_{\chi^0_1} h^2$ are also studied.

\section{Relic Density of the Lightest Neutralino}

The present mass density of the lightest neutralino 
is inversely proportional to the annihilation cross section 
of $\chi^0_1$ pairs.
When kinematically allowed, the neutralino pairs annihilate 
into fermion pairs ($f\bar{f}$), gauge boson pairs ($W^+W^-$ and $ZZ$), 
Higgs boson pairs\footnote{There are five Higgs bosons: 
two CP-even $h^0$ (lighter) and $H^0$ (heavier), 
one CP-odd $A^0$, and a pair of singly charged $H^\pm$.}  
($h^0h^0$, $H^0H^0$, $A^0A^0$, $h^0H^0$ $h^0A^0$, $H^0A^0$ and $H^+H^-$) 
and associated pairs of gauge and Higgs bosons 
($Zh^0$, $ZH^0$, $ZA^0$, and $W^\pm H^\mp$) 
through s, t, and u channel diagrams.
The most convenient approach to calculate the annihilation cross section 
including interferences among all diagrams is to calculate the amplitude 
of each diagram with the helicity amplitude formalism\footnote{
The helicity amplitudes of all relevant processes 
in the non-relativistic limit are given in Ref.~\cite{Manuel}.}
then numerically evaluate the matrix element squared with a computer.
Therefore, following Ref.~\cite{Howie}, 
we employ the HELAS package \cite{HELAS} to calculate the helicity amplitudes.
The annihilation cross section of $\chi^0_1$ pairs 
depends on its mass ($m_{\chi^0_1}$) which we find can be empirically 
expressed in GeV units as a function of the GUT scale gaugino mass $m_{1/2}$ 
and $\sin 2\beta$ as
\begin{eqnarray}
\mu > 0: &  & \;\; m_{\chi^0_1} \simeq 0.448 m_{1/2} +11.7 \sin2 \beta -10.4, 
                   \nonumber \\
\mu < 0: &  & \;\; m_{\chi^0_1} \simeq 0.452 m_{1/2} +4.68 \sin 2\beta -12.9,
\end{eqnarray}
for 100 GeV $\alt m_{1/2} \alt$ 1000 GeV and all $\tan\beta$ 
for which perturbative RGE solutions exist.
These neutralino mass formulas hold to an accuracy of $\alt 3\%$.

In the regions of Breit-Wigner resonance poles 
the near-singular amplitudes make the integration unstable.
To greatly improve the efficiency and accuracy in calculating 
the annihilation cross section for processes with a Breit-Wigner resonance, 
we introduce a transformation $s-M^2 = M\Gamma \tan\theta$ 
such that \cite{Vernon} 
\begin{equation}
\int \frac{ds}{ (s-M^2)^2+(M\Gamma)^2 } = \int \frac{d\theta}{M\Gamma}
\end{equation}
and scale $\theta$ to the range (0,1).

The time evolution of the number density $n(t)$ of weakly interacting 
mass particles is described by the Boltzmann equation \cite{Relic} 
\begin{equation}
\frac{dn}{dt} = -3 H n -\left< \sigma v \right>[ n^2 -n_{E}^2 ]
\end{equation}
where $H = 1.66 g_*^{1/2} T^2/M_{Pl}$ is the Hubble expansion rate 
with $g_* \simeq 81$ the effective number of relativistic degrees of freedom, 
$M_{Pl} = 1.22\times 10^{19}$ is the Planck mass,
$\left<\sigma v\right>$ is the thermally averaged cross section 
times velocity, 
$v$ is the relative velocity 
and $\sigma$ is the annihilation cross section, 
and $n_E$ is the number density at thermal equilibrium.

In the early Universe, when the temperature $T \gg m_{\chi^0_1}$, 
the LSP existed abundantly in thermal equilibrium 
with the LSP annihilations into lighter particles 
balanced by pair production.
Deviation from the thermal equilibrium began when the temperature reached 
the freeze-out temperature ($T_f \simeq m_{\chi^0_1}/20$). 
After the temperature dropped well below $T_f$, 
the annihilation rate became equal to the expansion rate   
and $n_{\chi^0_1} = H/\left<\sigma v\right>$. 
The resulting relic density can be expressed 
in terms of the critical density by \cite{Rocky}
\begin{equation}
\Omega_{\chi^0_1} h^2 = m_{\chi^0_1} n_{\chi^0_1}/\rho_c
\end{equation}
where $n_{\chi^0_1}$ is the number density of $\chi^0_1$. 

The thermally averaged cross section times velocity is 
\begin{eqnarray}
\left<\sigma v\right>(T)={
{\int (\sigma v) e^{-E_1/T} e^{-E_2/T} d^3p_1 d^3p_2}\over
{\int e^{-E_1/T} e^{-E_2/T} d^3p_1 d^3p_2} },
\end{eqnarray}
where $p_1$ ($E_1$) and $p_2$ ($E_2$) are the momentum and energy of the two
colliding particles in the cosmic, co-moving frame of reference, and $T$
is the temperature. This expression has been simplified 
to a one-dimensional integral \cite{Gondolo} of the form 
\begin{eqnarray}
\left<\sigma v\right> (x) = {1\over{4xK_2^2({1\over x})}}
\int_2^{\infty} da \sigma (a) a^2(a^2-4) K_1\left( {a\over x} \right),
\end{eqnarray}
where $x={T\over m_{\chi^0_1}}$, 
$a={\sqrt{s}\over m_{\chi^0_1}}$, 
$\sqrt{s}$ is the subprocess energy, 
and the $K_i$ are modified Bessel functions of order $i$.

To calculate the neutralino relic density, 
it is necessary to first determine the scaled freeze-out temperature 
$x_F \equiv m_{\chi^0_1}/T_f$.
The standard procedure is to iteratively solve the freeze out relation
\begin{eqnarray}
x_F^{-1} = \log\left[ \frac{ m_{\chi^0_1} }{ 2\pi^3 }
               \sqrt{ \frac{45}{2g_* G_N} } 
               \left<\sigma v\right>_{x_F} x_F^{1/2} \right],
\end{eqnarray}
starting with $x_F = \frac{1}{20}$. 
The relic density at the present temperature ($T_0$) is then obtained from
\begin{equation}
\Omega h^2= \frac{ \rho (T_0) }{ 8.0992\times 10^{-47} {\rm GeV}^4 },
\end{equation}
where
\begin{eqnarray}
\rho (T_0) \simeq 1.66\times \frac{1}{M_{Pl}}
\left(\frac{ T_{\chi^0_1} }{ T_\gamma }\right)^3
T_\gamma^3 {\sqrt{g_*}} \frac{1}{ \int_0^{x_F}
\left<\sigma v\right> dx }.
\label{eq:rho}
\end{eqnarray}
The ratio $T_{\chi^0_1}/T_\gamma$ is the reheating factor \cite{Reheating} 
of the neutralino temperature ($T_{\chi^0_1}$) 
compared to the microwave background temperature ($T_\gamma$).
The integration in Eq.~(\ref{eq:rho}) is evaluated 
with the modified Bessel functions expanded as power series in $x$, 
and then integrated over $x$.
The result is 
\begin{eqnarray}
\int_0^{x_F}\left<\sigma v\right> dx =
\frac{1}{ \sqrt{8\pi} } \int_2^\infty da \sigma (a) a^{3/2} (a^2-4) F(a),
\label{eq:sigmav}
\end{eqnarray}
where \cite{Howie} 
\begin{eqnarray}
F(a)
& = &\sqrt{{\pi\over {a-2}}} 
     \left\{ 1-Erf\left(\sqrt{{{a-2}\over x_F}}\right)\right\} 
     \nonumber \\
&   &+2\left({3\over 8a}-{15\over 4}\right)
       \left\{ \sqrt{x_F} e^{-{{a-2}\over x_F}} -\sqrt{\pi (a-2)} 
            \left[ 1-Erf\left(\sqrt{{{a-2}\over x_F}}\right) \right] \right\} 
     \nonumber \\
&   & +{2\over 3}
       \left( {285\over 32}-{45\over 32a}-{15\over 128a^2} \right)\times 
     \nonumber \\
&   &\left\{ e^{-{{a-2}\over x_F}}\left[ x_F^{3\over 2}-2(a-2)\sqrt{x_F}\right]
      +2\sqrt{\pi }(a-2)^{3\over 2}
       \left[ 1-Erf\left(\sqrt{{a-2}\over x_F} \right) \right] \right\}.
\end{eqnarray}
Essentially, all of the contribution to the integral in Eq.~(\ref{eq:sigmav}) 
comes from $a < 2.5$. 

To generate the SUSY particle mass spectrum and couplings 
for the neutralino relic density calculation, 
we run the gauge coupling RGEs from the weak scale 
up to a high energy scale\footnote{
For simplicity, we employed the one-loop RGEs in our analysis 
since the two-loop modifications to the SUSY mass spectra are only 
a few percent of the one-loop RGE results \cite{BBO}. 
We did not include heavy particle threshold effects in the RGE evolution, 
which may introduce small corrections 
when $\tan\beta$ is large \cite{Tomas}.}  
to fix the unification scale $M_{\rm GUT}$, 
and the unified gauge coupling $\alpha_{\rm GUT}$. 
Then we run all the RGEs from $M_{\rm GUT}$ to the weak scale of $M_Z$.
The SUSY mass scale ($M_{\rm SUSY}$) is chosen to be 2 TeV; 
closely similar results are obtained with a 1 TeV mass scale.
%

In Figure 1, we present masses, in the case\footnote{
Our convention for the sign of $\mu$ follows that of Ref. \cite{BBO} 
and is opposite to that of Ref. \cite{Howie}; 
see Eq. (\ref{eq:xino}).} $\mu > 0$, 
for the lightest neutralino ($\chi^0_1$), 
the lighter top squark ($\tilde{t}_1$),
the lighter tau slepton ($\tilde{\tau}_1$),
and two neutral Higgs bosons: the lighter CP-even ($h^0$) 
and the  CP-odd ($A^0$). 
To a good approximation, 
the mass of the lightest chargino ($\chi^+_1$) is about twice $m_{\chi^0_1}$ 
and the mass of the heavier CP-even Higgs boson ($H^0$) 
is close to $m_A$ \cite{BBO,SUGRA2}.
For $m_0 \sim 100$ GeV and $m_{1/2} \agt 400$ GeV, 
the mass of $\tilde{\tau}_1$ can become smaller than $m_{\chi^0_1}$ 
so such regions are theoretically excluded. 
Also shown in Fig.~1 are the regions that do not satisfy 
the following theoretical requirements: 
electroweak symmetry breaking (EWSB), tachyon free, 
and the lightest neutralino as the LSP.
The region excluded by the $m_{\chi^+_1} > 85$ GeV limit 
from the chargino search \cite{ALEPH} at LEP 2 is indicated.
Masses for $\mu < 0$ are shown in Fig. 2.

There are several interesting aspects to note in Figs. 1 and 2: \\
(i) For $\tan\beta \sim 1.8$, 
$\mu >0$ generates heavier neutralinos and charginos 
but lighter $h^0$ than $\mu < 0$ for the same $m_{1/2}$ and $m_0$. \\
(ii) An increase in $\tan\beta$ leads to a larger $m_h$ 
but a reduction in $m_{\chi^0_1}$, $m_{\chi^\pm_1}$, $m_A$ and $m_H$. \\
(iii) For $\tan\beta \sim 50$, $m_A \sim m_H$ can become comparable 
to $m_{\chi^0_1}$. \\ 
(iv) Increasing $m_0$ raises $m_A$, $m_H$ and the masses of other scalars 
significantly. 

The relic density of the neutralino dark matter 
($\Omega_{\chi^0_1} h^2$) for the case $\mu > 0$ 
is presented in Fig. 3 versus $m_{1/2}$ 
for several values of $m_0$ and $\tan\beta$. 
Figure 4 shows similar contours for $\mu < 0$.
When $2m_{\chi^0_1}$ is close to the mass of $Z$, $h^0$, $H^0$, 
or $A^0$, the cross section is significantly enhanced 
by the Breit-Wigner resonance pole, 
and a corresponding dip appears in the relic density distribution. 
For $\mu >0$, $\tan\beta = 1.8$ and $m_0 = 100$ GeV, there are two dips,
the first one occurs at the $h^0$ pole ($m_{1/2} = 55$ GeV, 
$m_{\chi^0_1} = 26$ GeV and $m_h = 52$ GeV), 
and the second one appears at the $Z$ pole ($m_{1/2} \simeq 100$ GeV 
and $m_{\chi^0_1} = 45$ GeV $\sim \frac{1}{2} M_Z$).
An increase in $m_0$ raises the mass of Higgs bosons 
but does not affect the mass of $\chi^0_1$.
Therefore, comparing the $m_0 = 100$ GeV results to those of $m_0 = 500$ GeV, 
the $Z$-pole dips occur at about the same $m_{1/2}$ as above, but the $h$-pole dips appear at larger $m_{1/2}$ 
and $m_{\chi^0_1}$.  
Increasing $\tan\beta$ slightly raises $m_h$, 
slightly lowers $m_{\chi^0_1}$, and greatly reduces $m_A$ and $m_H$. 
For $\tan\beta \sim 45$, extra dips appear in Figs. 3(c) and 4(c). 
These resonance dips are associated with the $A^0$ and $H^0$ masses 
and occur where $2m_{\chi^0_1} \simeq m_A \simeq m_H$.
For $\tan\beta \le 10$ and $m_0 > 200$ GeV, the annihilation cross section 
is too small for a cosmologically interesting $\Omega_{\chi^0_1} h^2$, 
while for $\tan\beta \sim 45$, 
$m_0$ must be close to 1000 GeV for $m_{1/2} \alt 750$ GeV and 
be larger than 400 GeV for $m_{1/2} \agt 750$ GeV 
to obtain an interesting relic density. 
For $\tan\beta \agt 50$, $m_A \alt 2m_{\chi_1^0}$, 
and there are no $A^0$ and $H^0$ resonant contributions; 
an acceptable $\Omega h^2$ is found for $m_0 \agt 400$ GeV 
and 400 GeV $\alt m_{1/2} \alt$ 800 GeV.

\section{Constraints on the SUGRA parameter space}

To show the constraints of Eq. (\ref{eq:ohs}) on the SUGRA parameter space, 
we present contours of $\Omega_{\chi^0_1} h^2 = 0.1$ and 0.5 in the 
($m_{1/2},\tan\beta$) plane in Fig. 5 for $\mu > 0$.
Also shown are the regions that do not satisfy 
the theoretical requirements (electroweak symmetry breaking, 
tachyon free, and lightest neutralino as the LSP) 
and the region excluded by the chargino search 
($m_{\chi^+_1} < 85$ GeV) at LEP 2 \cite{ALEPH}.
Figure 6 shows similar contours for $\mu < 0$.

We summarize below the central features of these results. \\
If $m_0$ is close to 100 GeV, 
\begin{itemize}
\item 
Most of the ($m_{1/2},\tan\beta$) plane
with $\tan\beta \agt 40$ is excluded by the above mentioned 
theoretical requirements.
\item 
The chargino search at LEP 2 excludes the region where $m_{1/2} \alt 100$ GeV
for $\mu > 0$ and $m_{1/2} \alt$ 120 GeV for $\mu < 0$.
\item
The cosmologically interesting region is 100 GeV $\alt m_{1/2} \alt$ 400 GeV 
and $\tan\beta \alt 25$ for either sign of $\mu$.
\end{itemize}
If $m_0$ is close to 500 GeV,
\begin{itemize}
\item 
Most of the ($m_{1/2},\tan\beta$) plane 
is theoretically acceptable for $\tan\beta \alt 50$ 
and $m_{1/2} \agt 120$ GeV.
\item 
The LEP 2 chargino search excludes 
(i)  $m_{1/2} \alt 140$ GeV for $\tan\beta \agt 10$, and 
(ii) $m_{1/2} \alt 80$ GeV ($\mu > 0$) 
 or $m_{1/2} \alt 120$ GeV ($\mu < 0$) for $\tan\beta \sim 1.8$.
\item 
The cosmologically interesting regions lie in two narrow bands.  
\end{itemize}

Contours of $\Omega_{\chi^0_1} h^2 = 0.1$ and 0.5 in the ($m_{1/2},m_0$) plane 
are presented in Fig. 7 for $\mu > 0$ with $\tan\beta = 1.8$ and $50$.
Also shown are the theoretically excluded regions. 
Figure 8 shows similar contours for $\mu < 0$.
Salient features of these results are summarized in the following. \\
If $\tan\beta$ is close to 1.8, 
\begin{itemize}
\item 
Most of the ($m_{1/2},m_0$) parameter space is theoretically acceptable.
\item 
The chargino search at LEP 2 excludes the region where $m_{1/2} \alt 80$ GeV
for $\mu > 0$ and $m_{1/2} \alt$ 110 GeV for $\mu < 0$.
\item
Most of the cosmologically interesting region 
is 80 GeV $\alt m_{1/2} \alt$ 450 GeV and $m_0 \alt 200$ GeV.
\end{itemize}
If $\tan\beta$ is close to 50, 
\begin{itemize}
\item 
The theoretically acceptable region in the ($m_{1/2},m_0$) plane 
is constrained to have $m_0 \agt 160$ GeV and $m_{1/2} \agt 150$ GeV 
for $\tan\beta \sim 50$. 
\item 
The LEP 2 chargino search excludes 
(i)  the region with $m_{1/2} \alt 125$ GeV for $\mu > 0$ or
(ii) the region with $m_{1/2} \alt 135$ GeV for $\mu < 0$, 
which is already inside the theoretically excluded region.
\item 
The cosmologically interesting region lies in a band with 
(i)  475 GeV $\alt m_{1/2} \alt$ 800 GeV for $\mu > 0$, or 
(ii) 500 GeV $\alt m_{1/2} \alt$ 840 GeV for $\mu < 0$, 
and $m_0 \agt 300$ GeV.
\end{itemize}

To correlate the acceptable values of neutralino relic density 
to the masses of SUSY particles at the weak scale, 
we present $\Omega_{\chi^0_1} h^2$ 
and representative SUSY mass spectra versus $m_{1/2}$ in Fig. 9, 
for $\mu > 0$ 
with (a) $\tan\beta = 1.8$, $m_0 = 150$ GeV 
and  (b) $\tan\beta = 50$, $m_0 = 600$ GeV.
Figure 10 shows similar contours for $\mu < 0$. 
Also shown are the theoretically excluded regions 
and the excluded region from the chargino search at LEP 2.

The neutralino relic density 
and representative SUSY mass spectrum versus $m_0$ are presented 
in Fig. 11, for $\mu > 0$ 
with (a) $\tan\beta = 1.8$, $m_{1/2} = 150$ GeV 
and  (b) $\tan\beta = 50$, $m_{1/2} = 500$ GeV.
Figure 12 shows similar contours for $\mu < 0$. 
Also shown are the theoretically excluded regions. 

The foregoing results were based on 
the GUT scale trilinear coupling choice $A_0 = 0$. 
To examine the sensitivity of the results to $A_0$, 
we present the neutralino relic density versus $A_0$ in Fig. 13 
for various values of $m_0 = m_{1/2}$ and $\tan\beta$.
For $\tan\beta \alt 10$, 
$\Omega_{\chi^0_1} h^2$ is almost independent of $A_0$. 
For $\tan\beta \sim 40$, 
the relic density is reduced by a positive $A_0$ 
but enhanced by a negative $A_0$.
Significant impact of $A_0$ on $\Omega_{\chi^0_1} h^2$ occurs 
only when both $m_0$ and $m_{1/2}$ are small and $\tan\beta$ is large.  
For $m_0 = m_{1/2} =$ 200 GeV and $\tan\beta = 40$ 
the theoretically allowed range of $A_0$ is 
$-380 \;\; {\rm GeV} \alt A_0 \alt 590 \;\; {\rm GeV}$. 
The value of $\Omega_{\chi^0_1} h^2$ with $A_0 =$ 590 GeV ($-$380 GeV) 
is about 0.22 (2.4) times the value with $A_0 =0$. 
For $m_0 = m_{1/2} =$ 800 GeV and $\tan\beta = 40$ 
the value of $\Omega_{\chi^0_1} h^2$ with $A_0 =$ 1000 GeV ($-$1000 GeV) 
is about 0.58 (1.7) times the value with $A_0 =0$. 

The dependence of $\Omega_{\chi^0_1} h^2$ on $A_0$ occurs 
because the masses of top squarks are sensitive to $A_t$, 
and the t-channel diagrams involving the stops make important contribution 
to neutralino annihilation cross section.  
In figures 14 ($\mu > 0$) and 15 ($\mu < 0$), we show the mass 
of the lighter top squark ($m_{\tilde{t}_1}$) 
and the  trilinear coupling $A_t$ 
versus $A_0$ for various values of $m_0 = m_{1/2}$ and $\tan\beta$.
The trilinear coupling $A_t$ is always negative and 
it is nearly a linear function of $A_0$.
For $\tan\beta$ close to the top Yukawa infrared fixed point, 
$\tan\beta \simeq 1.8$, 
$A_t$ approaches an infrared fixed point vaule as well, 
$A_t \simeq -1.7 m_{1/2}$, to about $30\%$ accuracy, 
independent of the value of $A_0$ \cite{Carena}.

\section{Conclusions}

The existence of dark matter in the Universe provides 
a potentially important bridge between particle physics and cosmology.
Within SUGRA GUTs, the lightest neutralino 
is the most likely dark matter candidate. 
We evaluated the neutralino relic density 
using helicity amplitude methods including all relevant diagrams. 
We applied a transformation to improve the efficiency and accuracy 
of annihilation cross section calculations 
in regions with Breit-Wigner resonance poles.
Requiring that the neutralino relic density should be in the 
cosmologically interesting region, we were able to place tight constraints  
on the SUGRA parameter space, especially in the plane of $m_{1/2}$ 
versus $\tan\beta$, since the mass of the lightest neutralino depends mainly 
on these two parameters. 
We derived empirical formulas relating $m_{\chi^0_1}$ 
to $m_{1/2}$ and $\sin\beta$. 
The cosmologically interesting regions of the parameter space 
with $\tan\beta$ close to the top Yukawa infrared fixed points are found to be 
\begin{eqnarray}
\tan\beta = 1.8: & & \;\;\;\,
80 \;\; {\rm GeV} \alt m_{1/2} \alt 450 \;\;{\rm GeV}
\;\; {\rm and} \;\; m_0 \alt 200 \;\; {\rm GeV} \nonumber \\
\tan\beta = 50:  & & \;\; 
500 \;\; {\rm GeV} \alt m_{1/2} \alt 800 \;\; {\rm GeV}
\;\; {\rm and} \;\; m_0 \agt 300 \;\; {\rm GeV} \nonumber
\end{eqnarray}
where the high $\tan\beta$ result is based on $A_0 =0$. 
Both regions are nearly independent of the sign of $\mu$.
We presented expectations for the SUSY particle mass spectra 
corresponding to these regions of parameter space. 
For the IRFP at $\tan\beta = 1.8$, 
the lighter scalar tau ($\tilde{\tau}_1$) 
and the lightest scalar neutrino ($\tilde{\nu}$) 
can be relatively light ($\alt 200$ GeV) 
in the cosmologically interesting region.

For $\tan\beta \alt 10$, the neutralino relic density $\Omega_{\chi^0_1} h^2$
is almost independent of the trilinear couplings $A_0$. 
For $\tan\beta \sim 40$, the relic density is reduced by a positive $A_0$ 
while enhanced by a negative $A_0$.
The value of $A_0$ significantly affects $\Omega_{\chi^0_1} h^2$ 
only when $\tan\beta$ is large and both $m_0$ and $m_{1/2}$ are small.  

Supernovae \cite{Supernova}, 
cosmic microwave background \cite{CMB} 
and other astronomical observations 
will soon provide precise information on the Hubble parameter $h$ 
and the cold dark matter component of $\Omega$. 
This information will pinpoint the SUGRA GUT parameters, 
which in turn predict the SUSY particle masses in these models. 

\newpage
\section*{Acknowledgments}

We are grateful to Howie Baer and Michal Brhlik for providing us 
with their computer code, 
and to Mike Berger, Gary Steigman and Dieter Zeppenfeld 
for valuable discussions. 
We thank Manuel Drees for beneficial comments 
regarding the choice of scale for the one-loop effective potential. 
This research was supported 
in part by the U.S. Department of Energy under Grant No. DE-FG02-95ER40896, 
and in part by the University of Wisconsin Research Committee 
with funds granted by the Wisconsin Alumni Research Foundation.

%


\begin{figure}
\centering\leavevmode
\epsfysize=6.4in\epsffile{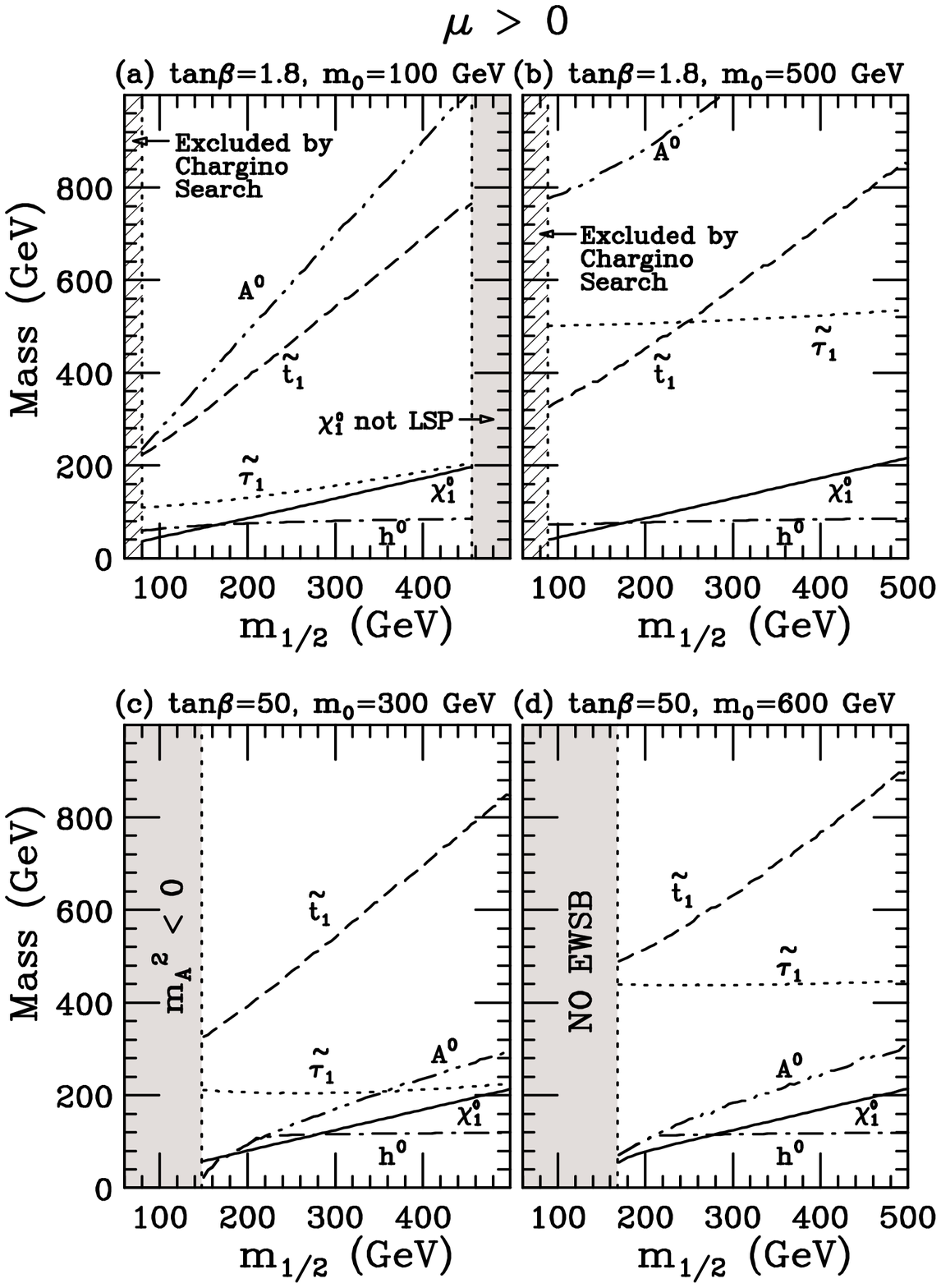}
\bigskip

\caption[]{
Masses of $\chi^0_1$, $\tilde{t}_1$, $\tilde{\tau}_1$ 
at the mass scale of $M_Z$ and 
masses of $h^0$, and $A^0$ at the mass scale 
$Q = \sqrt{  m_{\tilde{t}_L} m_{\tilde{t}_R} }$, 
versus $m_{1/2}$,  
with $M_{\rm SUSY} =$ 2 TeV and $\mu >0$ for 
(a)~$\tan\beta = 1.8$, $m_0 = 100$ GeV, 
(b)~$\tan\beta = 1.8$, $m_0 = 500$ GeV, 
(c)~$\tan\beta = 50$, $m_0 = 300$ GeV, and 
(d)~$\tan\beta = 50$, $m_0 = 600$ GeV. 
The shaded regions 
are excluded by theoretical requirements (EWSB, tachyons, LSP), 
or the chargino search at LEP 2.
\label{fig:mass1}
}\end{figure}

\begin{figure}
\centering\leavevmode
\epsfysize=6.4in\epsffile{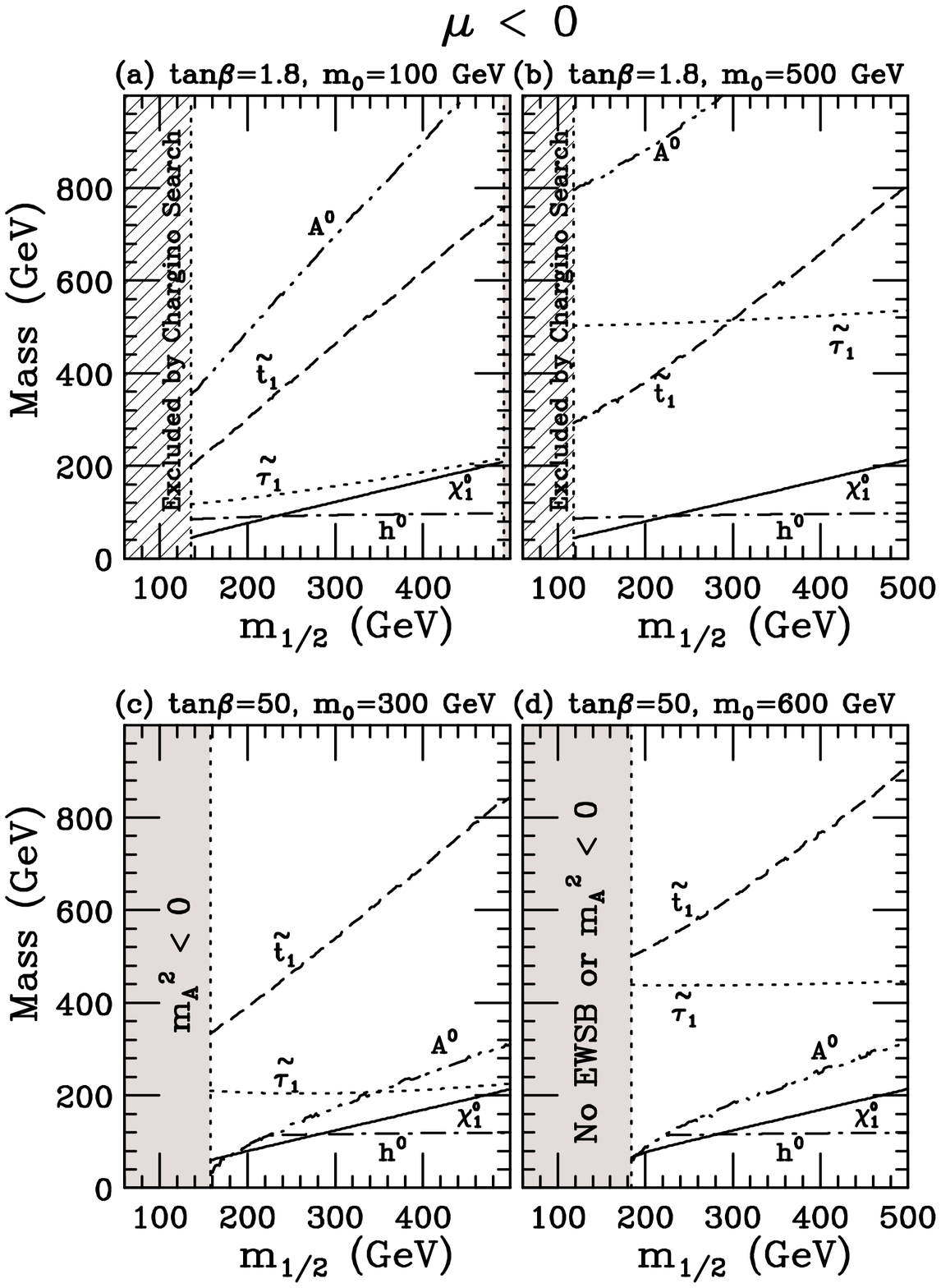}
\bigskip

\caption[]{
The same as in Fig. 1, except that $\mu < 0$.
\label{fig:mass2}
}\end{figure}

\begin{figure}
\centering\leavevmode
\epsfysize=6.4in\epsffile{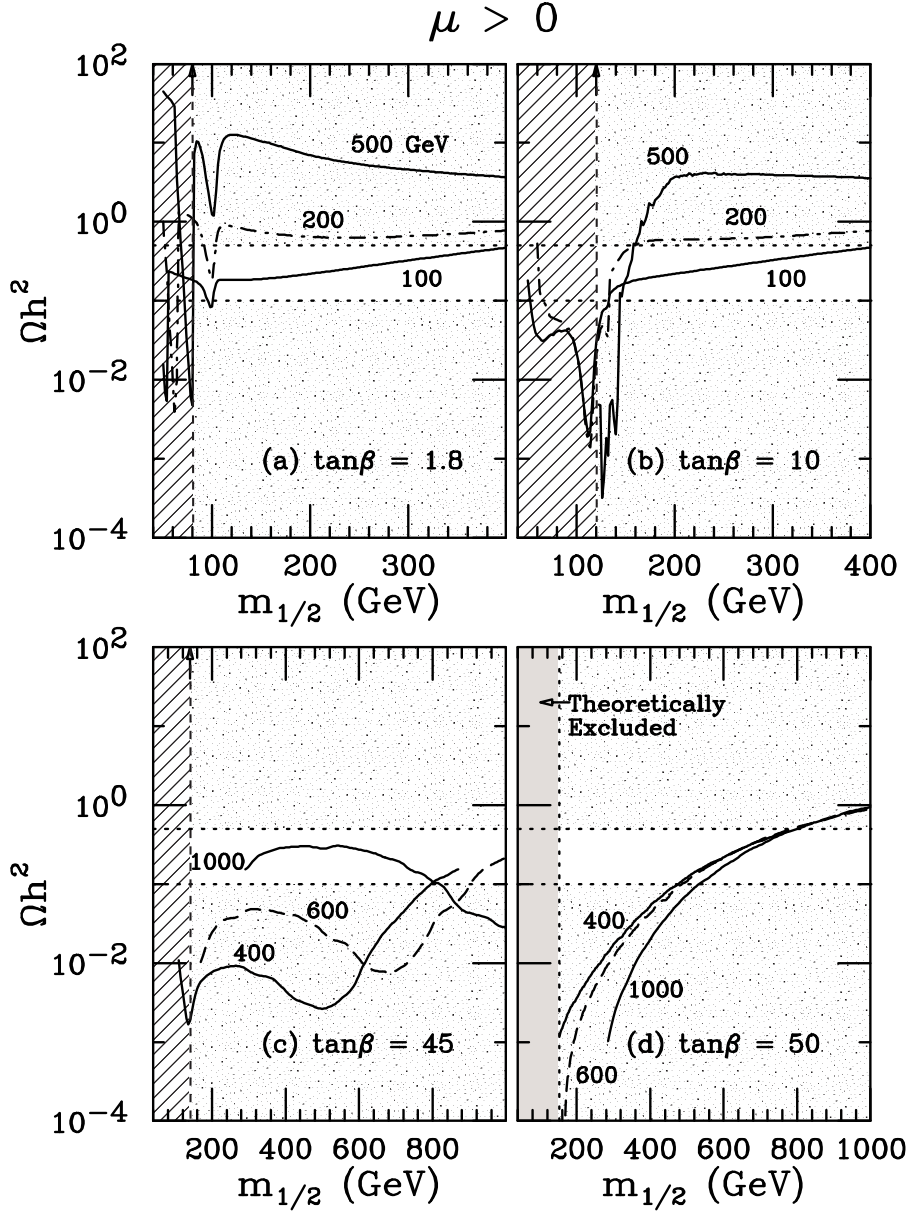}
\bigskip

\caption[]{
$\Omega_{\chi^0_1} h^2$ versus $m_{1/2}$ for $\mu >0$, 
with representative values of $m_0$ and 
(a)~$\tan\beta = 1.8$, 
(b)~$\tan\beta = 10$, 
(c)~$\tan\beta = 45$, and 
(d)~$\tan\beta = 50$.
The shaded regions denote the parts of the parameter space 
(i)~producing $\Omega_{\chi^0_1} h^2 < 0.1$ or $\Omega_{\chi^0_1} h^2 > 0.5$, 
(ii)~excluded by theoretical requirements, 
or (iii) excluded by the chargino search at LEP 2.
\label{fig:ohs1}
}\end{figure}

\begin{figure}
\centering\leavevmode
\epsfysize=6.4in\epsffile{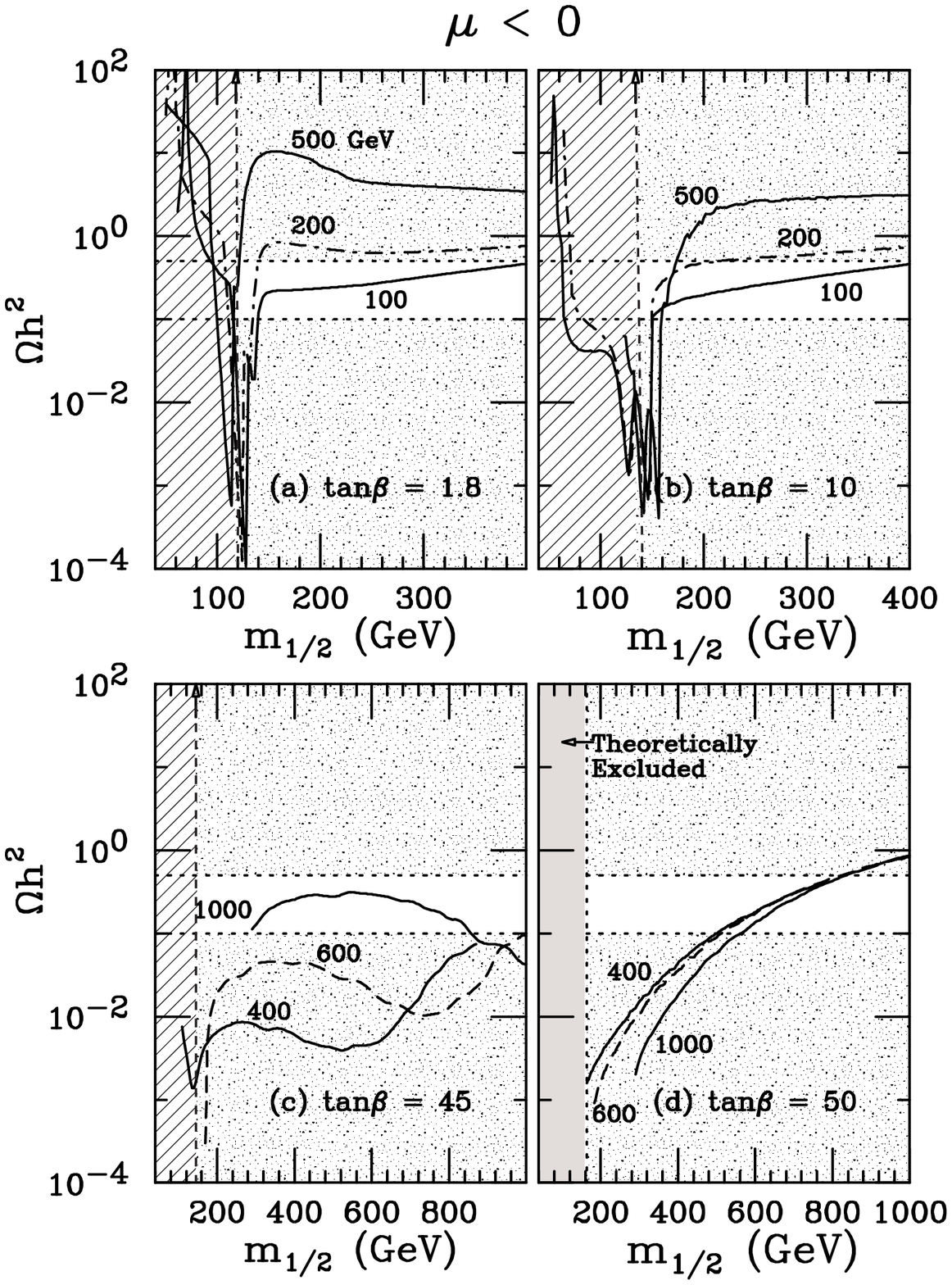}
\bigskip

\caption[]{
The same as in Fig.~3, except that $\mu < 0$.
\label{fig:ohs2}
}\end{figure}
%

\begin{figure}
\centering\leavevmode
\epsfxsize=6.4in\epsffile{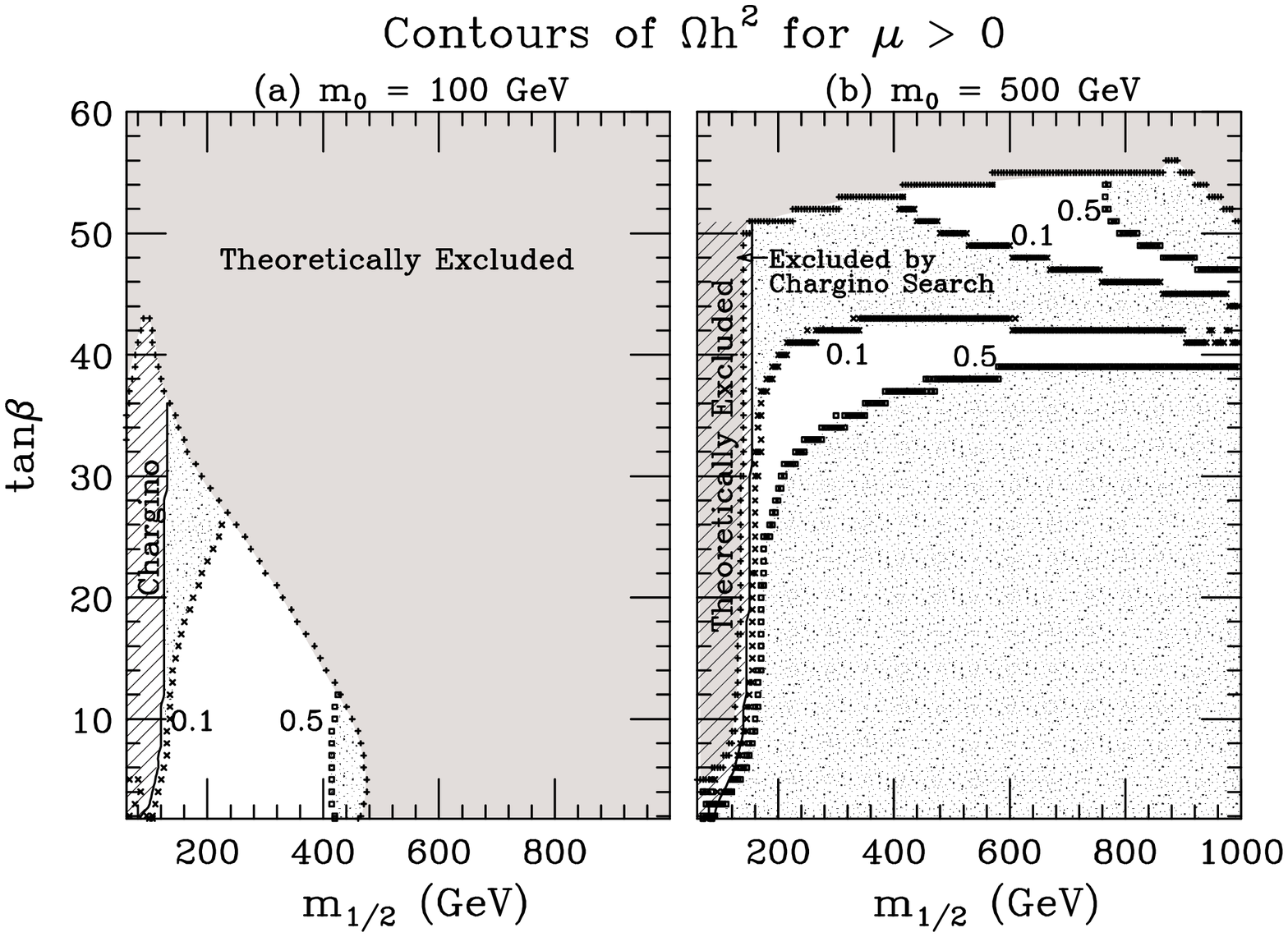}

\caption[]{
Contours of $\Omega_{\chi^0_1} h^2 = 0.1$ and 0.5 
in the $(m_{1/2},\tan\beta$) plane, 
for $\mu > 0$, (a) $m_0 = 100$ GeV and (b) $m_0 = 500$ GeV. 
The shaded regions denote the parts of the parameter space 
(i)~producing $\Omega_{\chi^0_1} h^2 < 0.1$ or $\Omega_{\chi^0_1} h^2 > 0.5$, 
(ii)~excluded by theoretical requirements, 
or (iii)~excluded by the chargino search at LEP 2.
\label{fig:xohs}
}\end{figure}
%

\begin{figure}
\centering\leavevmode
\epsfxsize=6.4in\epsffile{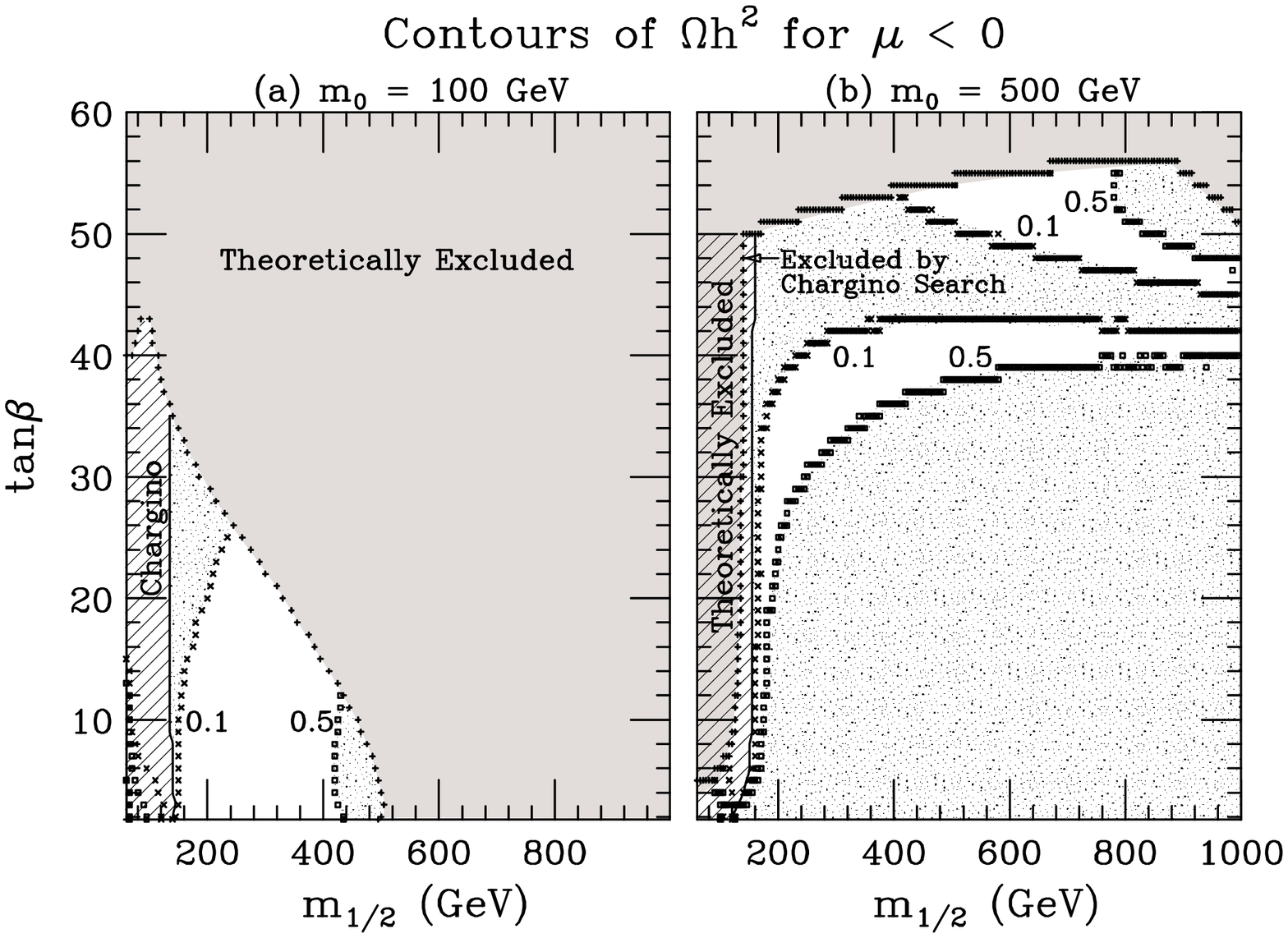}

\caption[]{
The same as in Fig. 5, except that $\mu < 0$. 
\label{fig:xohs2}
}\end{figure}
%

\begin{figure}
\centering\leavevmode
\epsfxsize=6.4in\epsffile{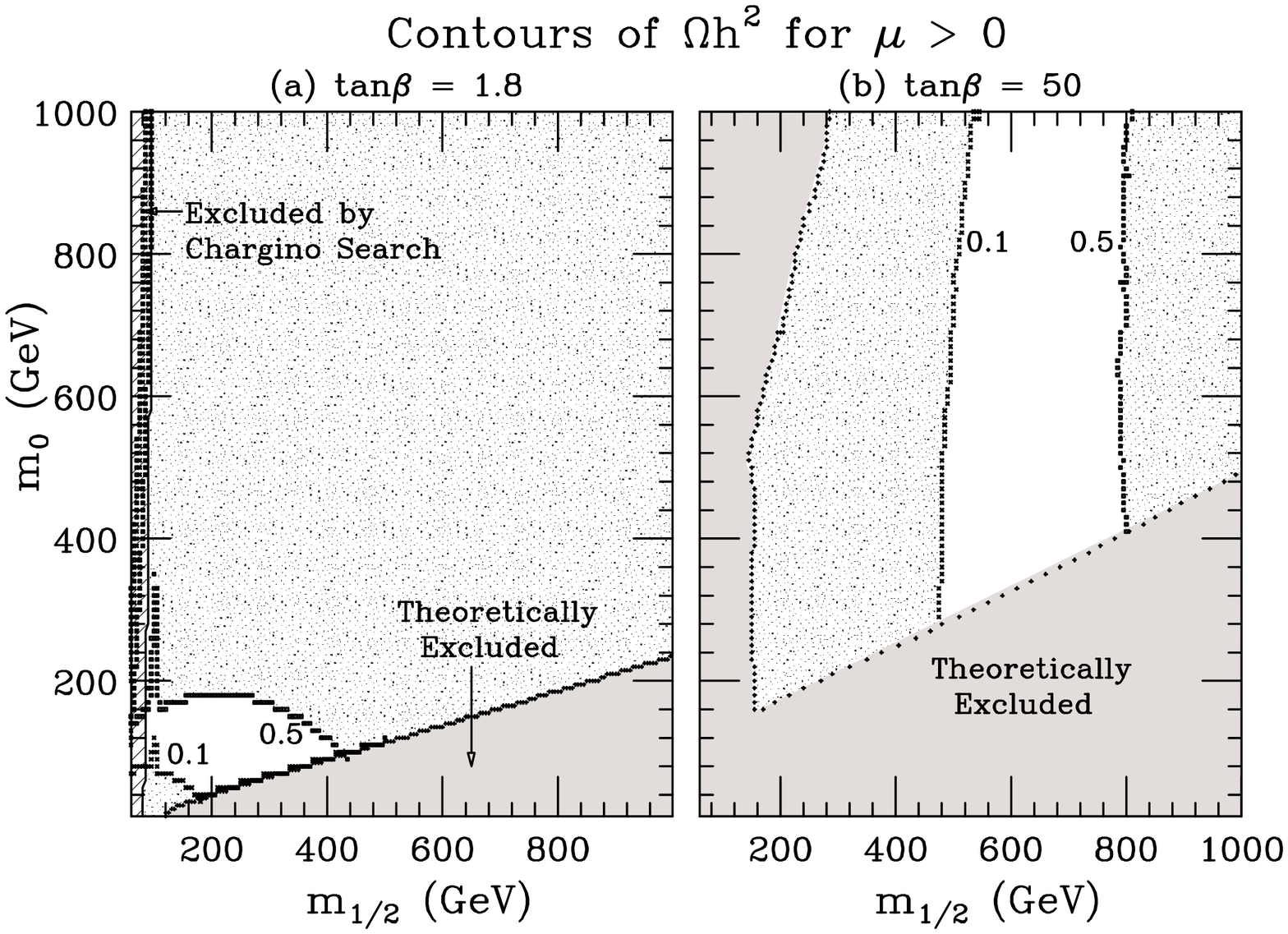}

\caption[]{
Contours of $\Omega_{\chi^0_1} h^2 = 0.1$ and 0.5 
in the $(m_{1/2},m_0$) plane, 
for $\mu > 0$, (a) $\tan\beta = 1.8$ and (b) $\tan\beta = 50$.
The shaded regions denote the parts of the parameter space 
(i)~producing $\Omega_{\chi^0_1} h^2 < 0.1$ or $\Omega_{\chi^0_1} h^2 > 0.5$, 
(ii)~excluded by theoretical requirements, 
or (iii)~excluded by the chargino search at LEP 2.
\label{fig:xohs3}
}\end{figure}
%

\begin{figure}
\centering\leavevmode
\epsfxsize=6.4in\epsffile{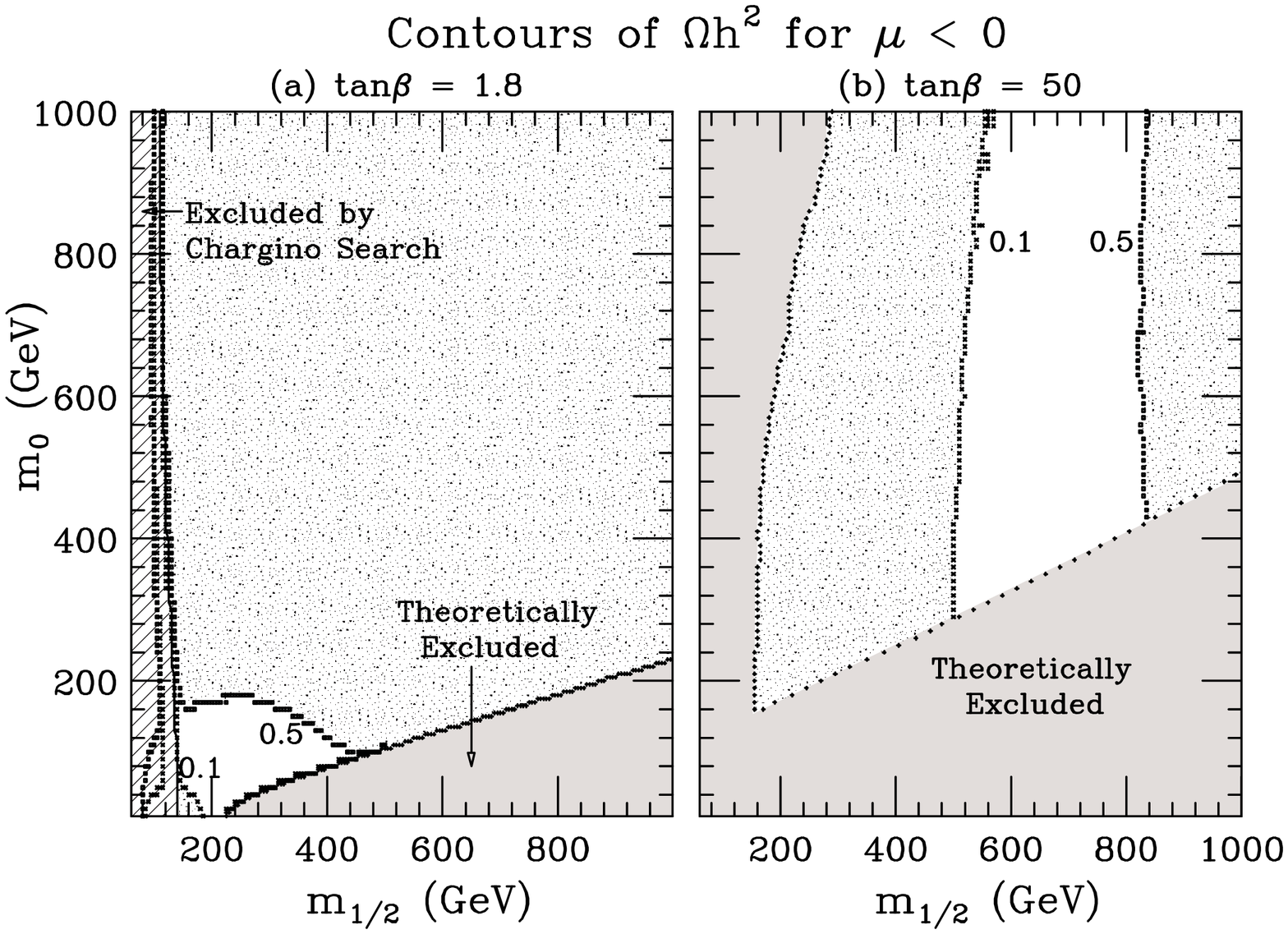}

\caption[]{
The same as in Fig. 7, except that $\mu < 0$.
\label{fig:xohs4}
}\end{figure}
%

\begin{figure}
\centering\leavevmode
\epsfxsize=6.4in\epsffile{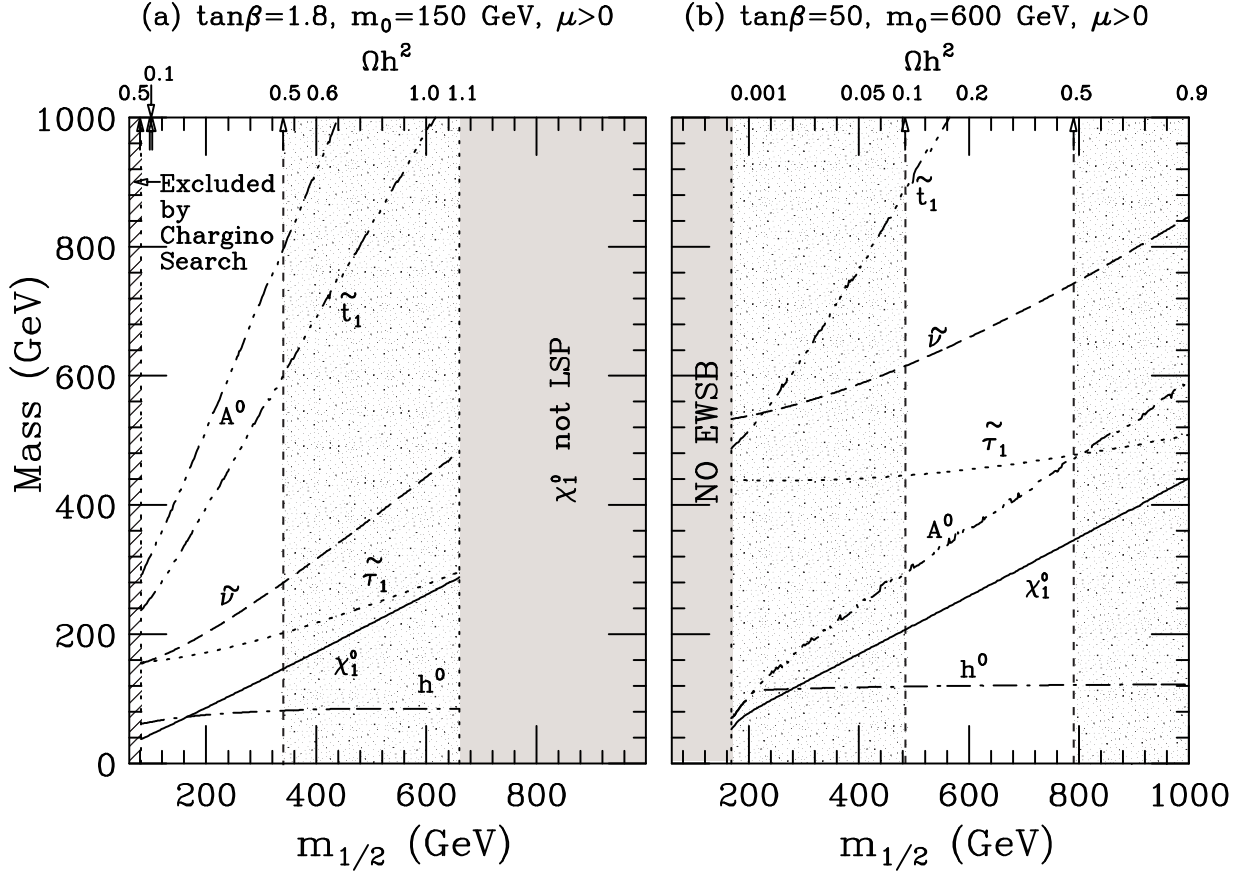}

\caption[]{
The neutralino relic density 
and representative SUSY mass spectrum versus $m_{1/2}$ for $\mu > 0$ 
with (a)~$\tan\beta = 1.8$, $m_0 = 150$ GeV and  
(b)~$\tan\beta = 50$, $m_0 = 600$ GeV.
In Figs. 9-12, $\tilde{\nu}$ is the lightest scalar neutrino.
The shaded regions denote the parts of the parameter space 
(i)~producing $\Omega_{\chi^0_1} h^2 < 0.1$ or $\Omega_{\chi^0_1} h^2 > 0.5$, 
(ii)~excluded by theoretical requirements, 
or (iii)~excluded by the chargino search at LEP 2.
\label{fig:mass9}
}\end{figure}
%

\begin{figure}
\centering\leavevmode
\epsfxsize=6.4in\epsffile{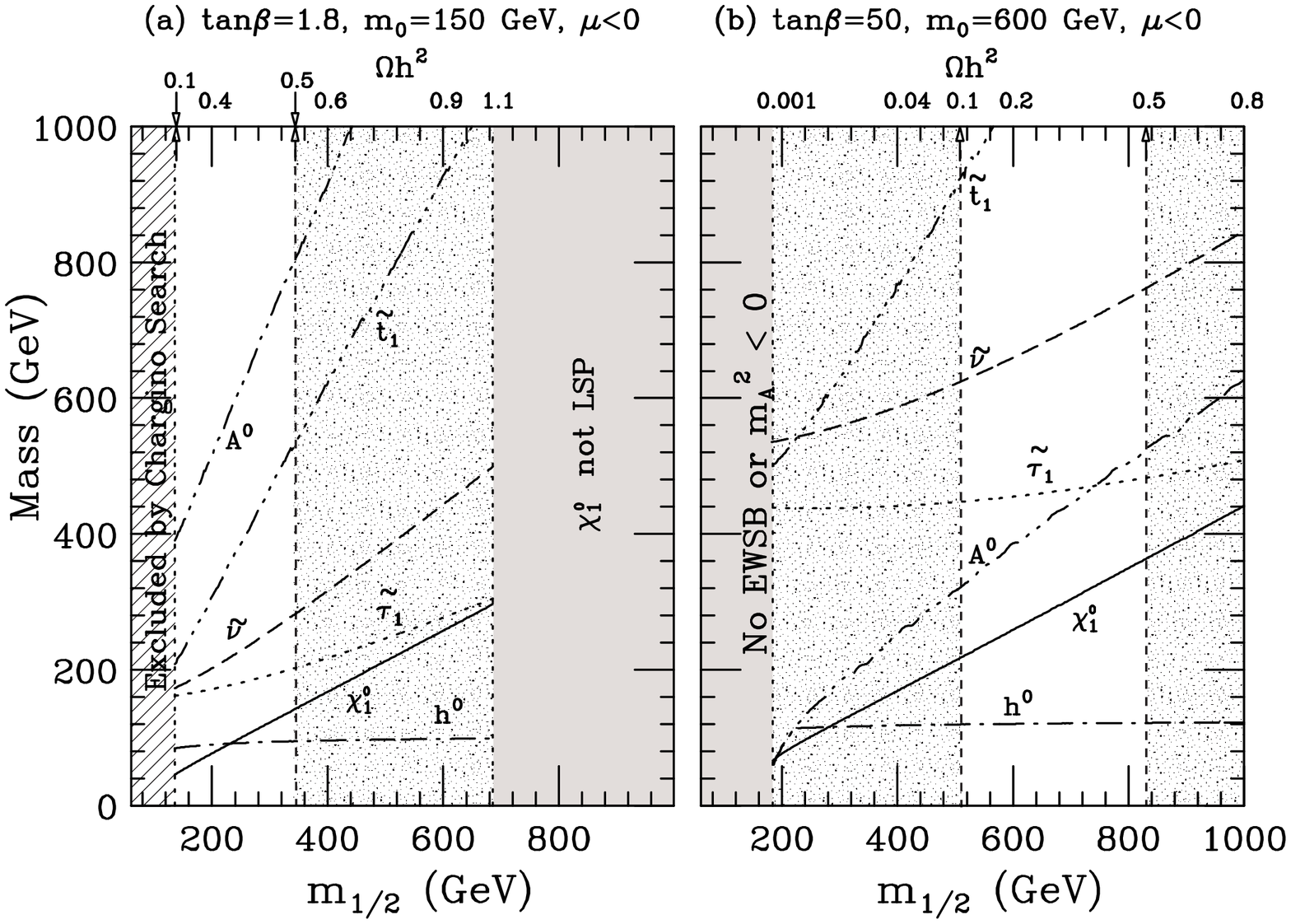}

\caption[]{
The same as in Fig. 9, except that $\mu < 0$. 
\label{fig:mass10}
}\end{figure}
%

\begin{figure}
\centering\leavevmode
\epsfxsize=6.4in\epsffile{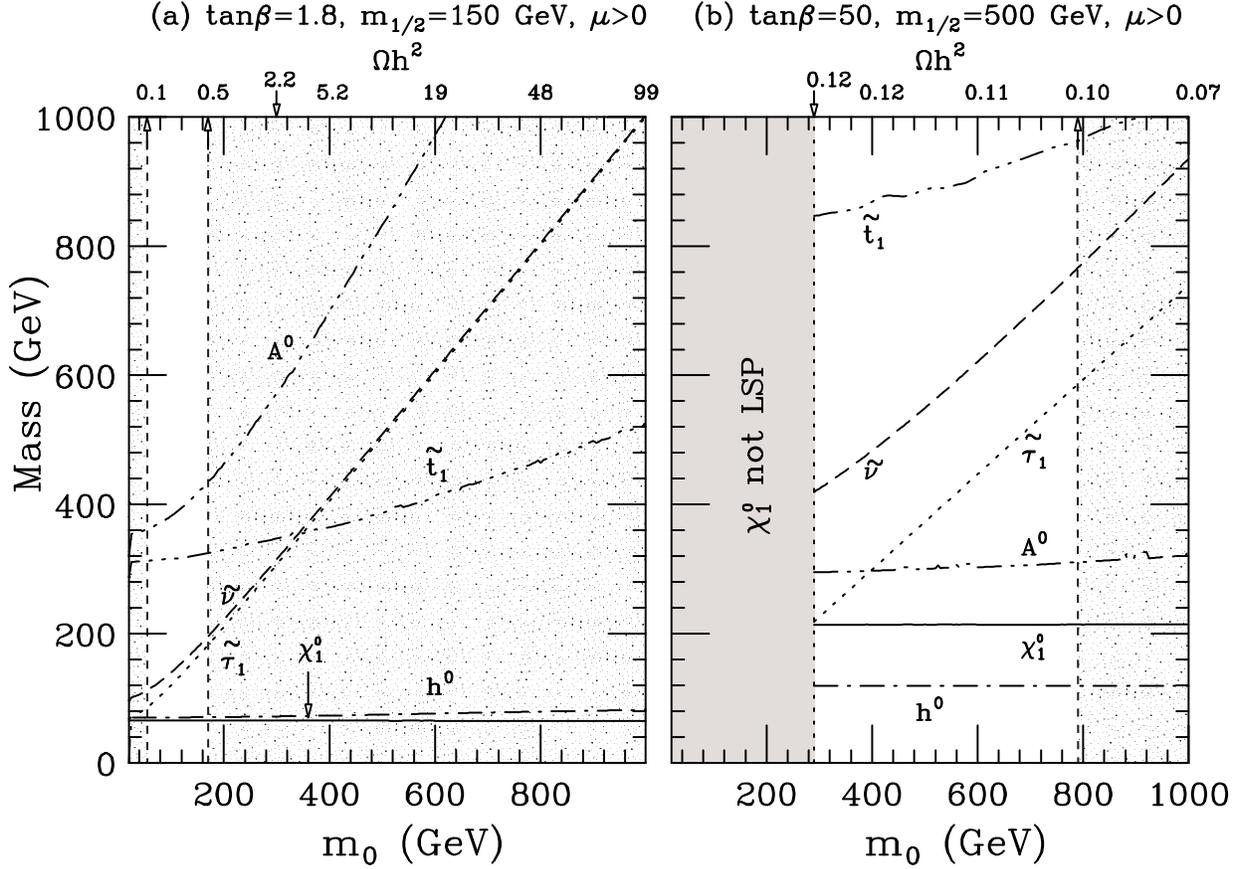}

\caption[]{
The neutralino relic density 
and representative SUSY mass spectrum versus $m_0$ for $\mu > 0$ 
with (a)~$\tan\beta = 1.8$, $m_{1/2} = 150$ GeV and 
(b)~$\tan\beta = 50$, $m_{1/2} = 500$ GeV.
The shaded regions denote the parts of the parameter space 
(i)~producing $\Omega_{\chi^0_1} h^2 < 0.1$ or $\Omega_{\chi^0_1} h^2 > 0.5$, 
or (ii)~excluded by theoretical requirements. 
\label{fig:mass11}
}\end{figure}
%

\begin{figure}
\centering\leavevmode
\epsfxsize=6.4in\epsffile{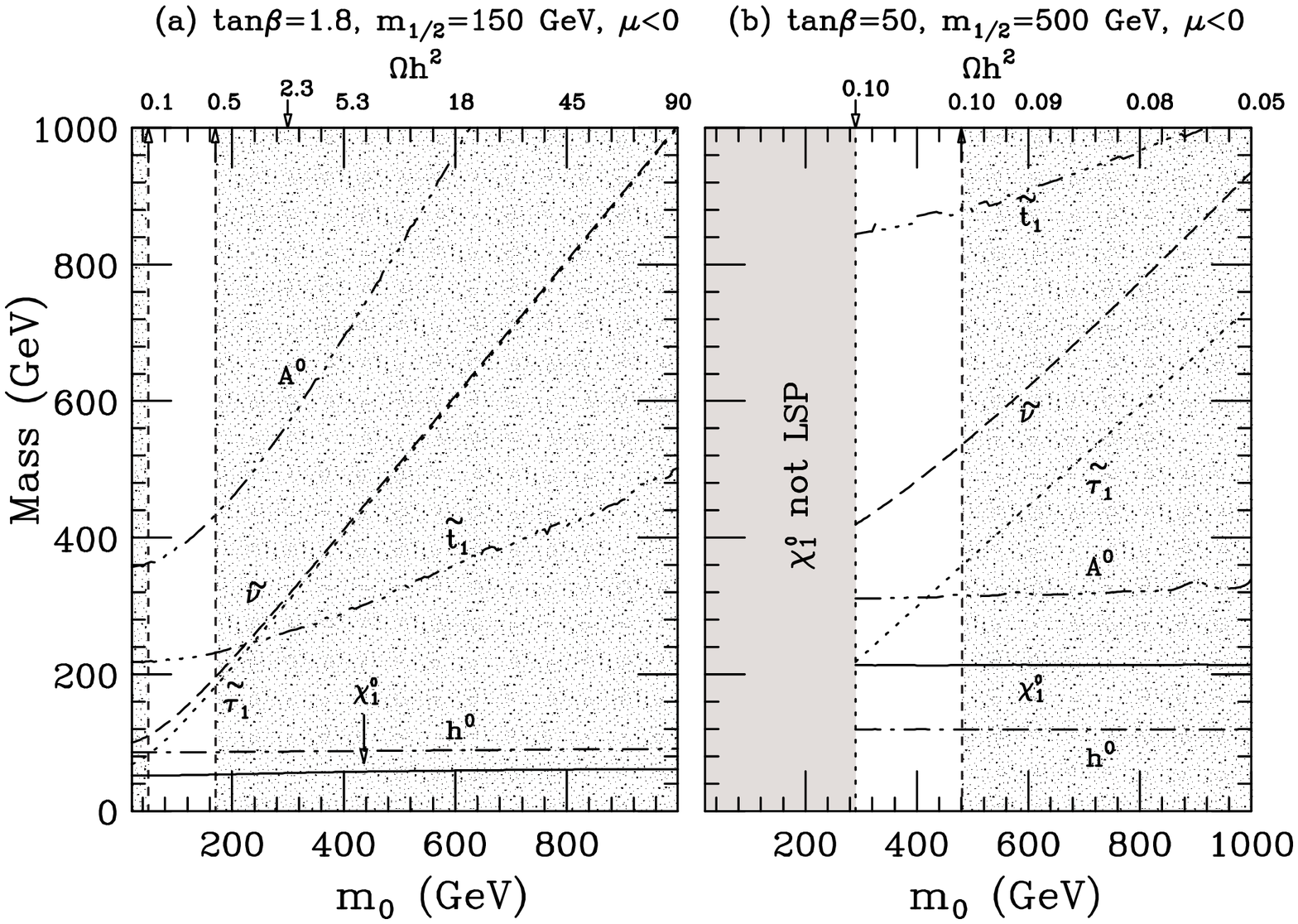}

\caption[]{
The same as in Fig. 11, except that $\mu < 0$.
\label{fig:mass12}
}\end{figure}
%

\begin{figure}
\centering\leavevmode
\epsfysize=7in\epsffile{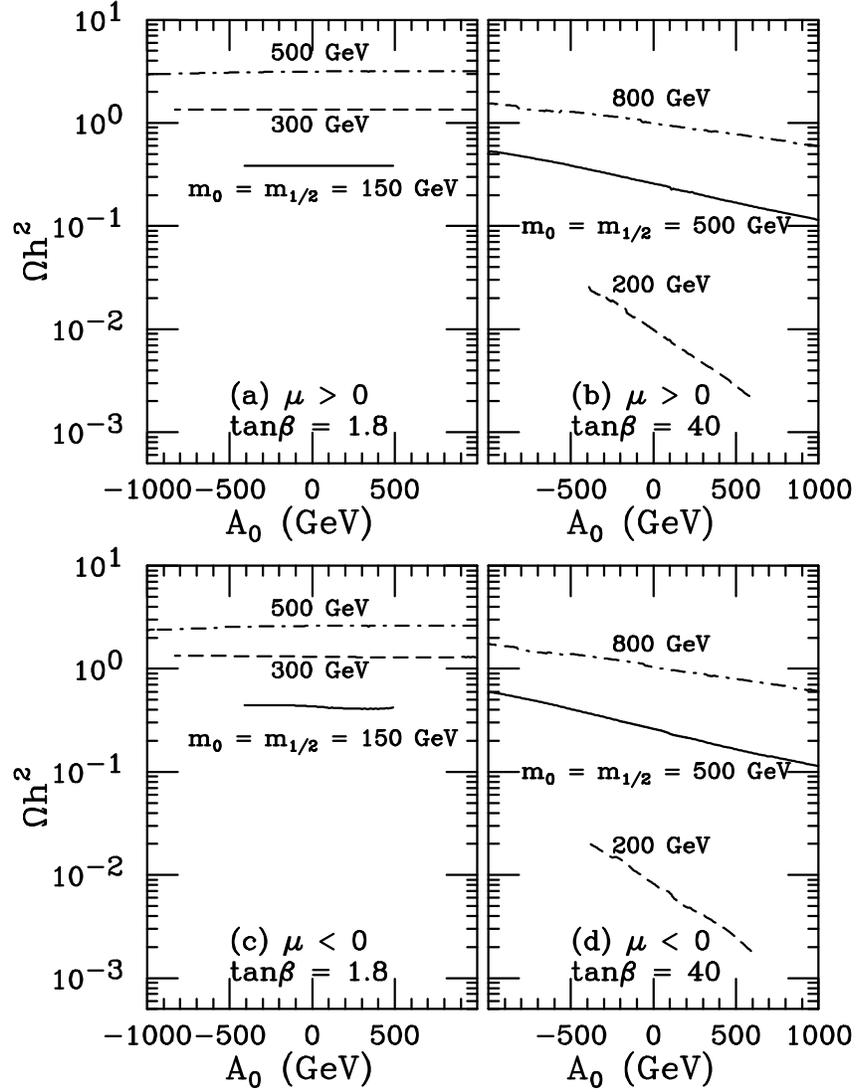}

\caption[]{
The neutralino relic density versus the GUT scalar trilinear coupling $A_0$ 
for 
(a) $\mu > 0$ and $\tan\beta = 1.8$, 
(b) $\mu > 0$ and $\tan\beta = 40$, 
(c) $\mu < 0$ and $\tan\beta = 1.8$, 
(d) $\mu < 0$ and $\tan\beta = 40$, 
and various values of $m_0 = m_{1/2}$. 
\label{fig:oha0}
}\end{figure}
%

\begin{figure}
\centering\leavevmode
\epsfysize=7in\epsffile{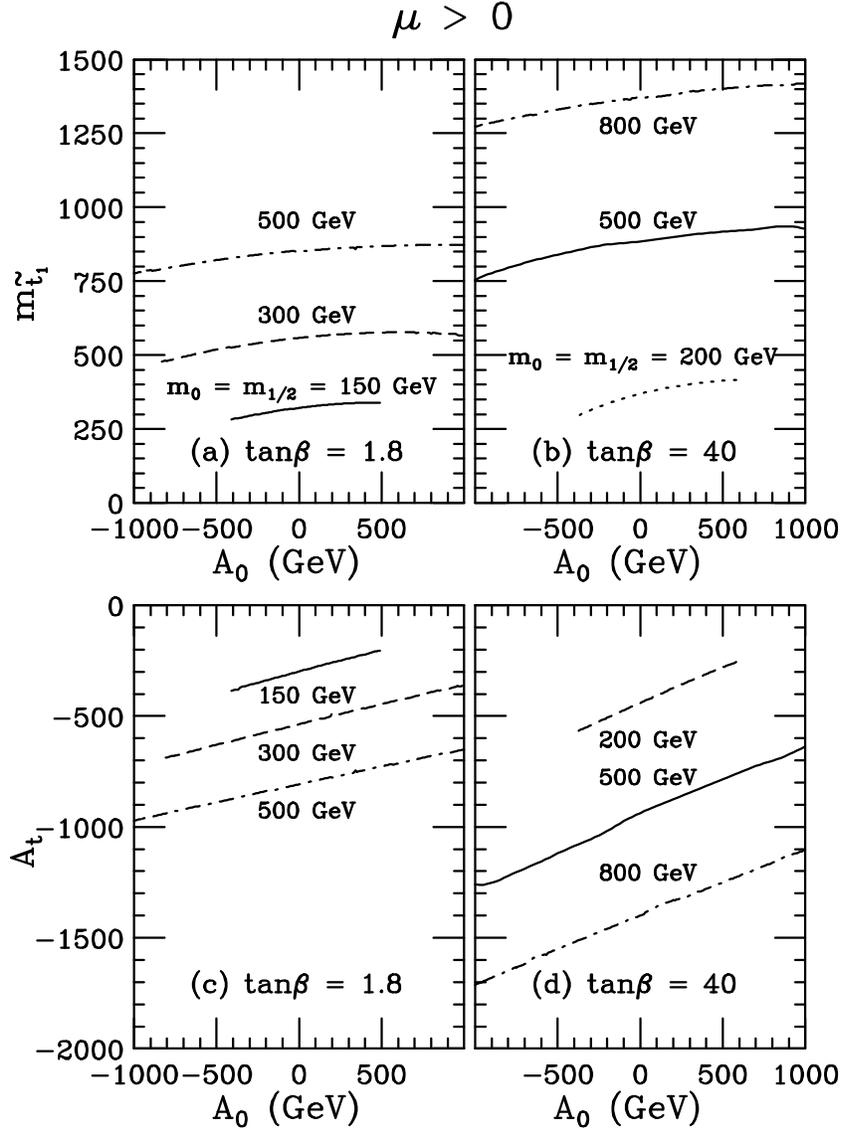}

\caption[]{
The mass of the lighter top squark versus $A_0$ for 
(a) $\tan\beta = 1.8$, 
(b) $\tan\beta = 40$, 
and various values of $m_0 = m_{1/2}$ and $\mu > 0$. 
Also shown is the soft breaking trilinear coupling $A_t$ at the weak scale 
versus the GUT scale input $A_0$ for 
(c) $\tan\beta = 1.8$, 
(d) $\tan\beta = 40$, 
\label{fig:mt1a}
}\end{figure}
%

\begin{figure}
\centering\leavevmode
\epsfysize=7in\epsffile{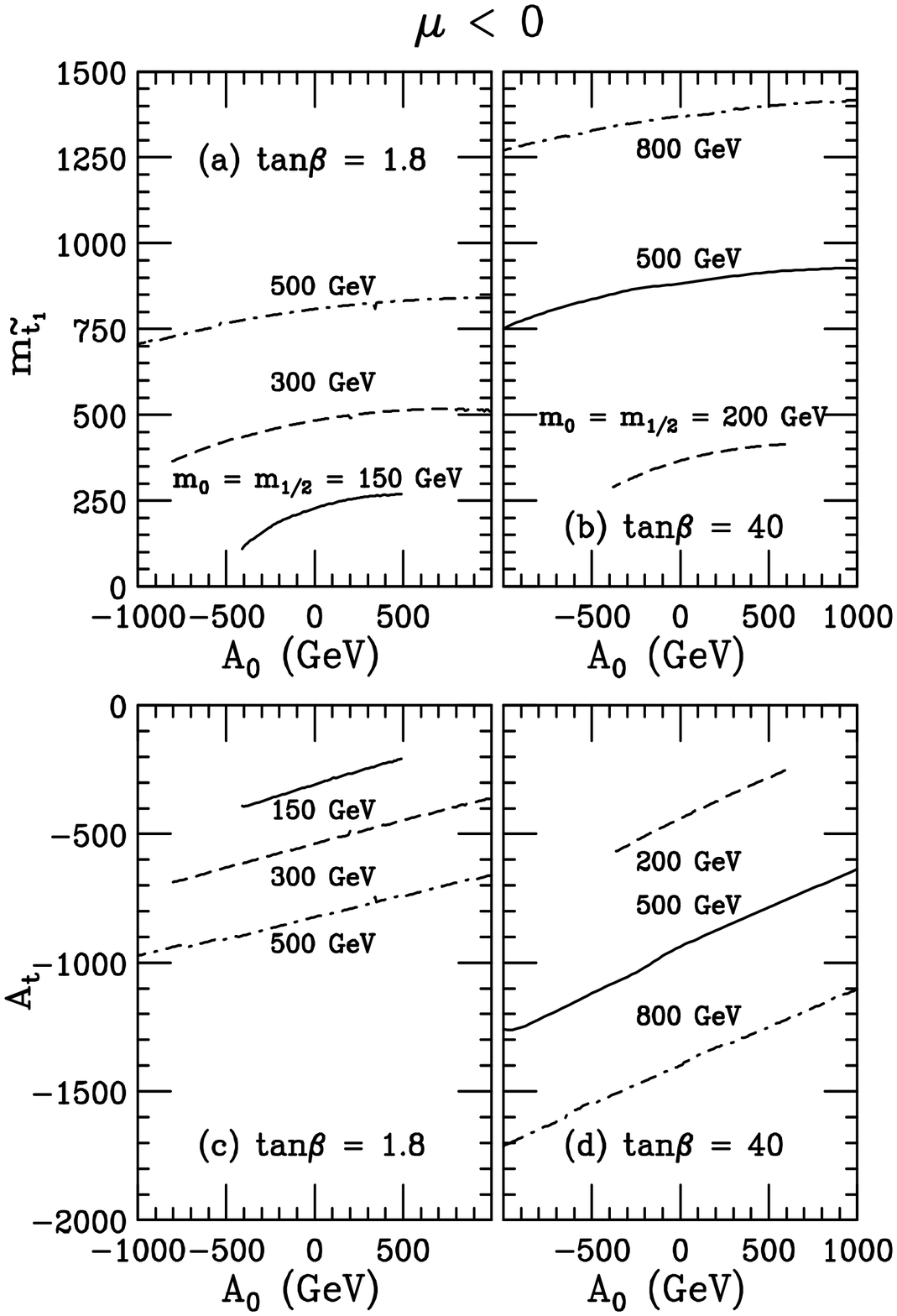}

\caption[]{
The same as in Fig. 14, except that $\mu < 0$.
\label{fig:mt1b}
}\end{figure}
%

\end{document}